\documentclass[
 reprint,
 amsmath,amssymb,
 aps,pra,
 groupedaddress,superscriptaddress,
 floatfix
]{revtex4-1}

\usepackage{graphicx}% Include figure files
\usepackage{dcolumn}% Align table columns on decimal point
\usepackage{bm}% bold maths

\usepackage{hyperref}% add hypertext capabilities
\hypersetup{
  colorlinks=true,
  citecolor=blue,
}

\usepackage[separate-uncertainty=true,multi-part-units=single]{siunitx}
\usepackage{braket}
\usepackage[version=4]{mhchem}

\begin{document}

%%%% FRONTMATTER %%%%
\title{Simultaneous Wide-field Imaging of Phase and Magnitude of AC Magnetic Signal Using Diamond Quantum Magnetometry}

\newcommand{\affTitech}{Department of Electrical and Electronic Engineering, School of Engineering, Tokyo Institute of Technology, 2-12-1 Ookayama, Meguro, Tokyo 152-8552, Japan}
\newcommand{\affPRESTO}{PRESTO, Japan Science and Technology Agency, 7 Gobancho, Chiyoda, Tokyo 152-0076, Japan}
\newcommand{\affQST}{National Institutes for Quantum and Radiological Science and Technology, 1233 Watanuki, Takasaki, Gunma 370-1292, Japan}

\author{Kosuke Mizuno}
  \email{mizuno.k.ae@m.titech.ac.jp}
  \affiliation{\affTitech}
\author{Hitoshi Ishiwata}
  \affiliation{\affTitech}
  \affiliation{\affPRESTO}
\author{Yuta Masuyama}
  \affiliation{\affQST}
\author{Takayuki Iwasaki}
  \affiliation{\affTitech}
\author{Mutsuko Hatano}
  \email{hatano.m.ab@m.titech.ac.jp}
  \affiliation{\affTitech}

\date{\today}

%%%% MACROS %%%%
\newcommand{\Ampere}{Amp\`{e}re}
\newcommand{\Koehler}{K\"{o}hler}
\newcommand{\Cramer}{Cram\`{e}r}

%%%% ABSTRACT %%%%

\begin{abstract}
Spectroscopic analysis of AC magnetic signal using diamond quantum magnetometry is a promising technique for inductive imaging.
Conventional dynamic decoupling like XY8 provides a high sensitivity of an oscillating magnetic signal with intricate dependence on magnitude and phase, complicating high throughput detection of each parameter.
In this study, a simple measurement scheme for independent and simultaneous detection of magnitude and phase is demonstrated by a sequential measurement protocol.
Wide-field imaging experiment was performed for an oscillating magnetic field with approximately \SI{100}{\micro m}-squared observation area.
Single pixel phase precision was \SI{2.1}{\degree} for \SI{0.76}{\micro T} AC magnetic signal.
Our method enables potential applications including inductive inspection and impedance imaging.
\end{abstract}

\maketitle

%%%% INTRODUCTION %%%%
\section{\label{sec:intro}Introduction}
A negatively charged nitrogen-vacancy (NV) center in diamond offers a promising material platform for quantum sensing~\cite{Degen2017,Casola2018}.
Spin-state manipulation with a state-selective microwave (MW) pulse, combined with spin-dependent fluorescence has been utilized for spectroscopic measurement of magnetic field.
Fabrication of a dense ensemble of NV centers~\cite{Pezzagna2010,Ozawa2019,Ishiwata2017,Tetienne2017a} allows these measurements to be applied in wide-field imaging modality~\cite{Ziem2019,Steinert2013,Horsley2018,Fescenko2018}, and local current characterizations using NV center have been achieved~\cite{Tetienne2017,Schlussel2018,Lillie2019,Tetienne2019,Zhou2019,Chatzidrosos2019}.
Wide-field imaging using NV center paves the way for inductive inspection~\cite{Flanagan1990,Ribeiro2008}, which is still challenging with micron-scale resolution by established methods.
Previous studies indicate that dynamical decoupling (DD) protocols like XY8 achieve magnetic field spectroscopy with high sensitivity for magnitude and phase of such a signal~\cite{Wolf2015,Barry2016,Masuyama2018,Ishikawa2019a}.
Despite its high sensitivity, DD can only measure the output of fluorescence intensity with an intricate dependence on a magnitude and a phase of a signal.
Therefore, a magnitude with a known phase or a phase with a known magnitude can only be measured with DD protocol~\cite{DeLange2011}.
Independent and simultaneous measurements of magnitude and phase for magnetic field spectroscopy in wide-field is required for an accurate inductive sensing.

In this study, we propose a stroboscopic measurement termed iQdyne~\cite{Mizuno2018}, a wide-field modality of Qdyne~\cite{Schmitt2017,Boss2017} enabled with lock-in detection, as a simple sensing scheme for magnitude $b_z$ and phase $\phi_0$ of an oscillating magnetic field.
The iQdyne provides an orthogonal measurement for magnitude and phase; it involves two input parameters $(b_z, \phi_0)$ and time-series outputs $I(b_z) e^{i \theta (\phi_0)} \cos⁡ (2\pi f t)$.
Fourier analysis easily extracts two resulting parameters, magnitude $I(b_z)$ and phase $\theta(\phi_0)$, which are separated from each other and are readily interpretable. 

We implemented this stroboscopic protocol on a wide-field microscope and demonstrated an imaging experiment of an oscillating magnetic field generated from a current pattern fabricated on a diamond substrate.
An oscillating current generates an oscillating magnetic field due to \Ampere'{}s law.
The wide-field observation area was approximately $100\times\SI{100}{{\micro m}^2}$, and we estimated that single pixel precision for phase sensing was \SI{2.1}{\degree} with \SI{0.76}{\micro T} AC magnetic signal.
This demonstration is a fundamental part of a local current investigation technique like local impedance microscopy and inductive testing method.
Wide-field modality provides us an optically high spatial resolution and a wide observation area, leading to high throughputs for such measurements.

%%%% FIGURES %%%%
\begin{figure*}
  \centering
  \includegraphics[width=13cm]{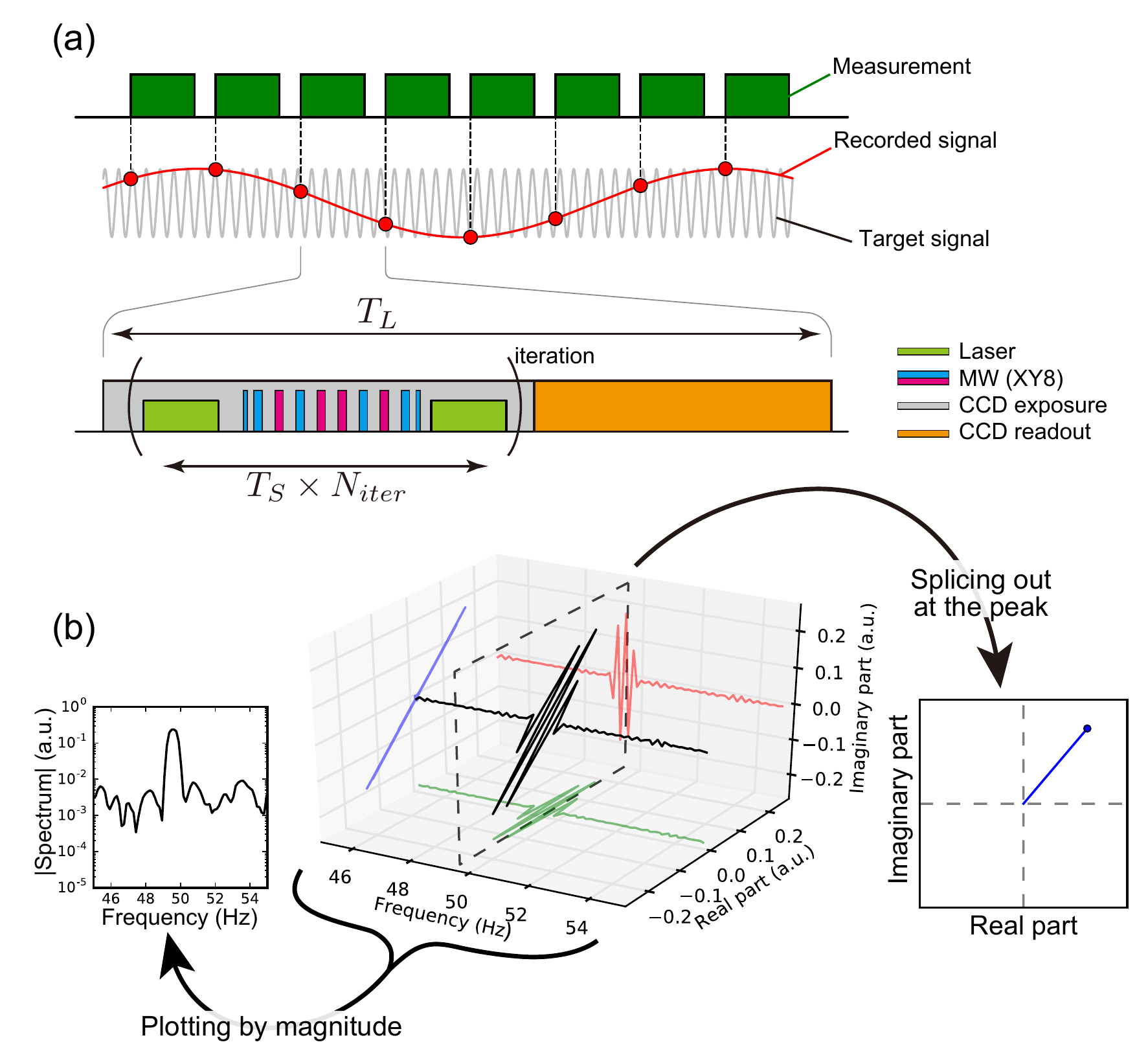}
  \caption{\label{fig1}(a)~Measurement protocol with sequential measurements recording the signal in under-sampling.  Each green represents quantum sensing iterated by $N_\mathrm{iter}$ times and a CCD readout.  (b)~Principle of an oscillating field analysis involving the three-dimensional Fourier spectrum: frequency axis, real, and imaginary part of the Fourier coefficient.  Splicing out at the signal frequency and moving to the IQ diagram, the Fourier spectrum represents the magnitude and the phase of the oscillating signal.}
\end{figure*}
\begin{figure}[htp]
  \centering
  \includegraphics[width=8.6cm]{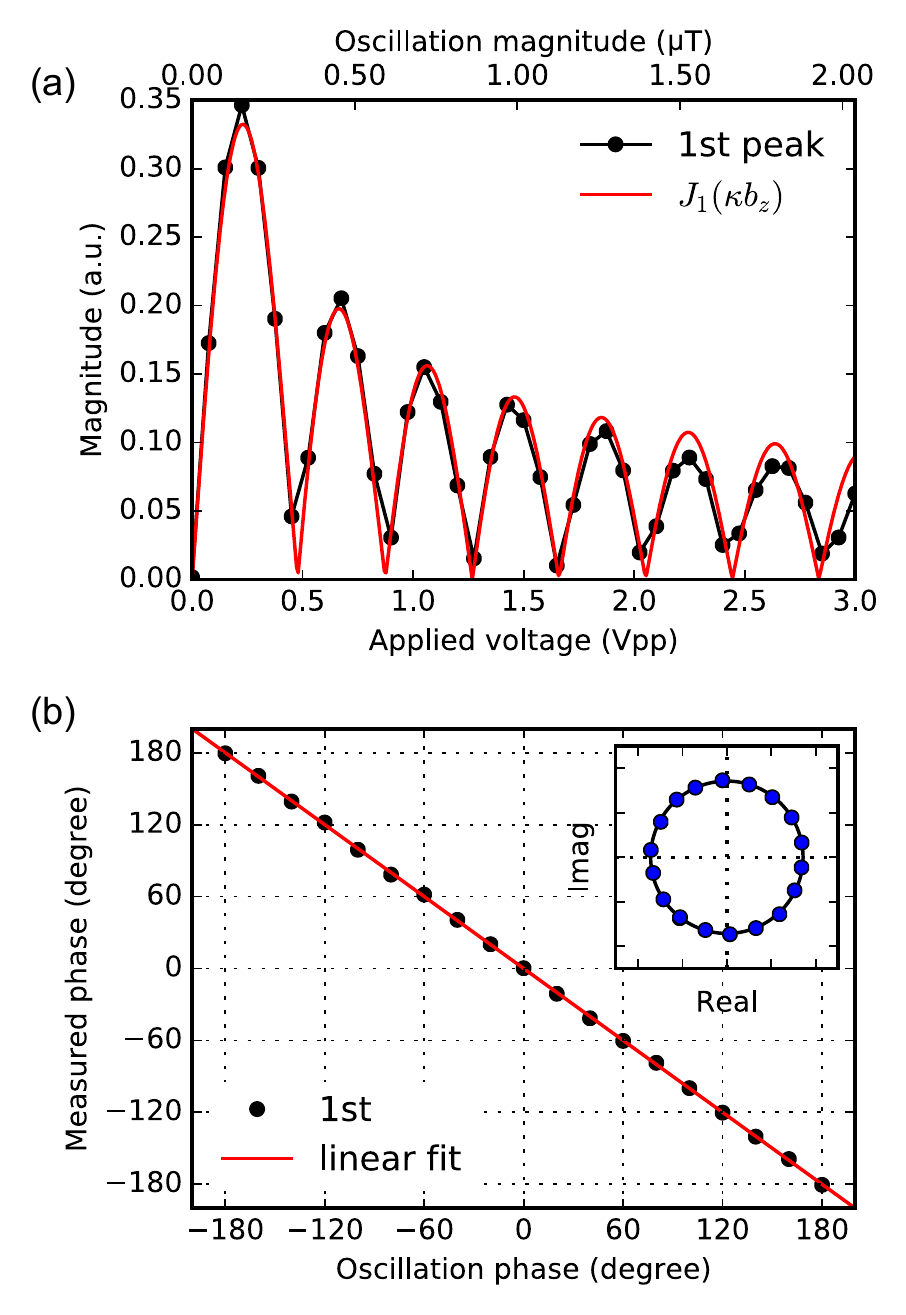}
  \caption{\label{fig2}The response of iQdyne-based measurement for an oscillating magnetic field; (a)~the magnitude of the NV center fluorescence of the first peak obeys the first-kind Bessel function with the oscillation field strength and (b)~the measured phase corresponds one-to-one to the oscillation phase.  Inset: plotted on an IQ diagram. }
\end{figure}
\begin{figure*}
  \centering
  \includegraphics[width=17.2cm]{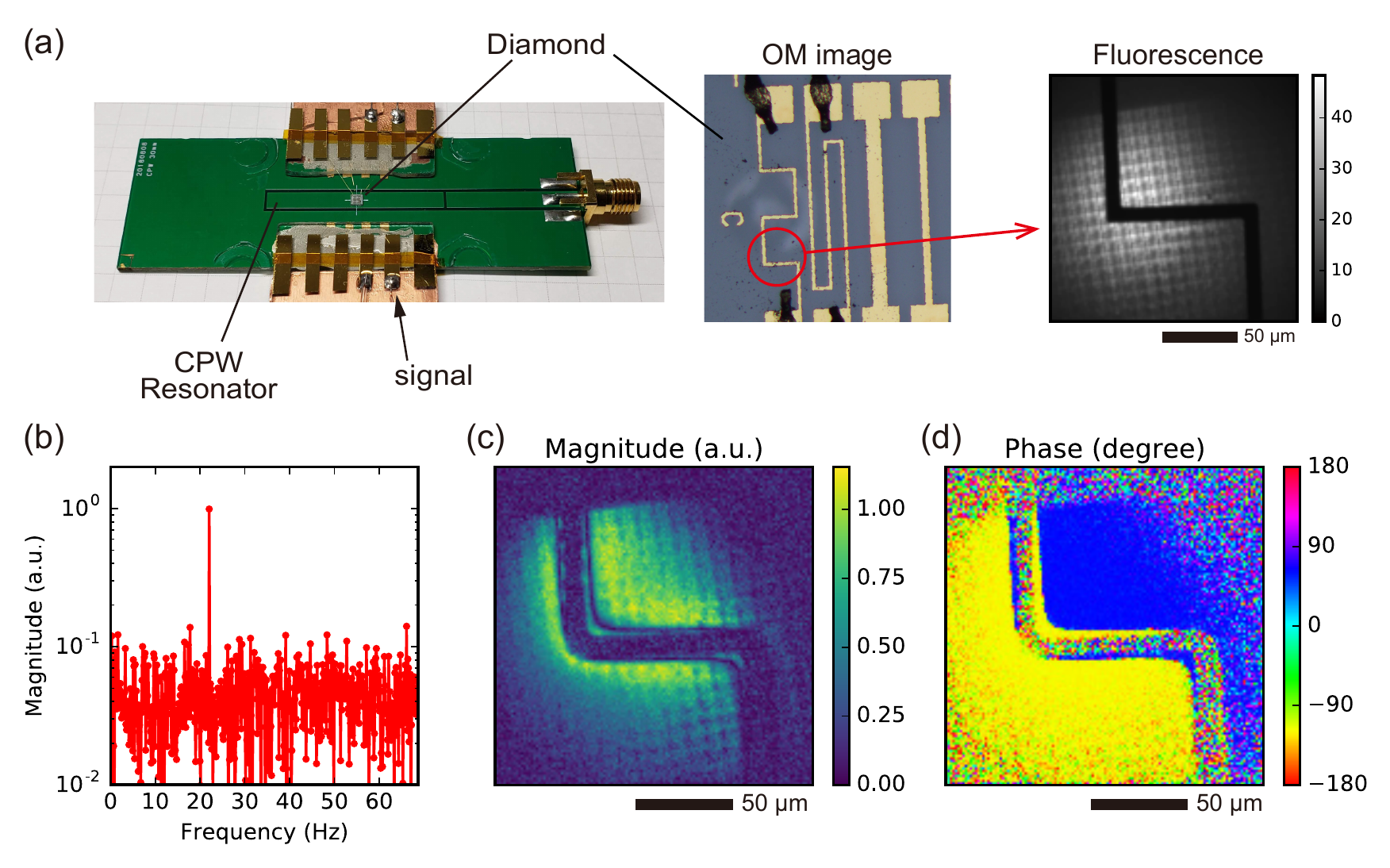}
  \caption{\label{fig3}Simultaneous wide-field imaging experiment involving (a)~the sample setup with the MW resonator to the left, the optical image of the current path as the signal source in the center, and fluorescence image to the right, (b)~typical measured spectrum showing frequency versus magnitude, (c)~the magnitude map, and (d)~phase map of the first peak at each pixel.}
\end{figure*}
\begin{figure}[htp]
  \centering
  \includegraphics[width=6.5cm]{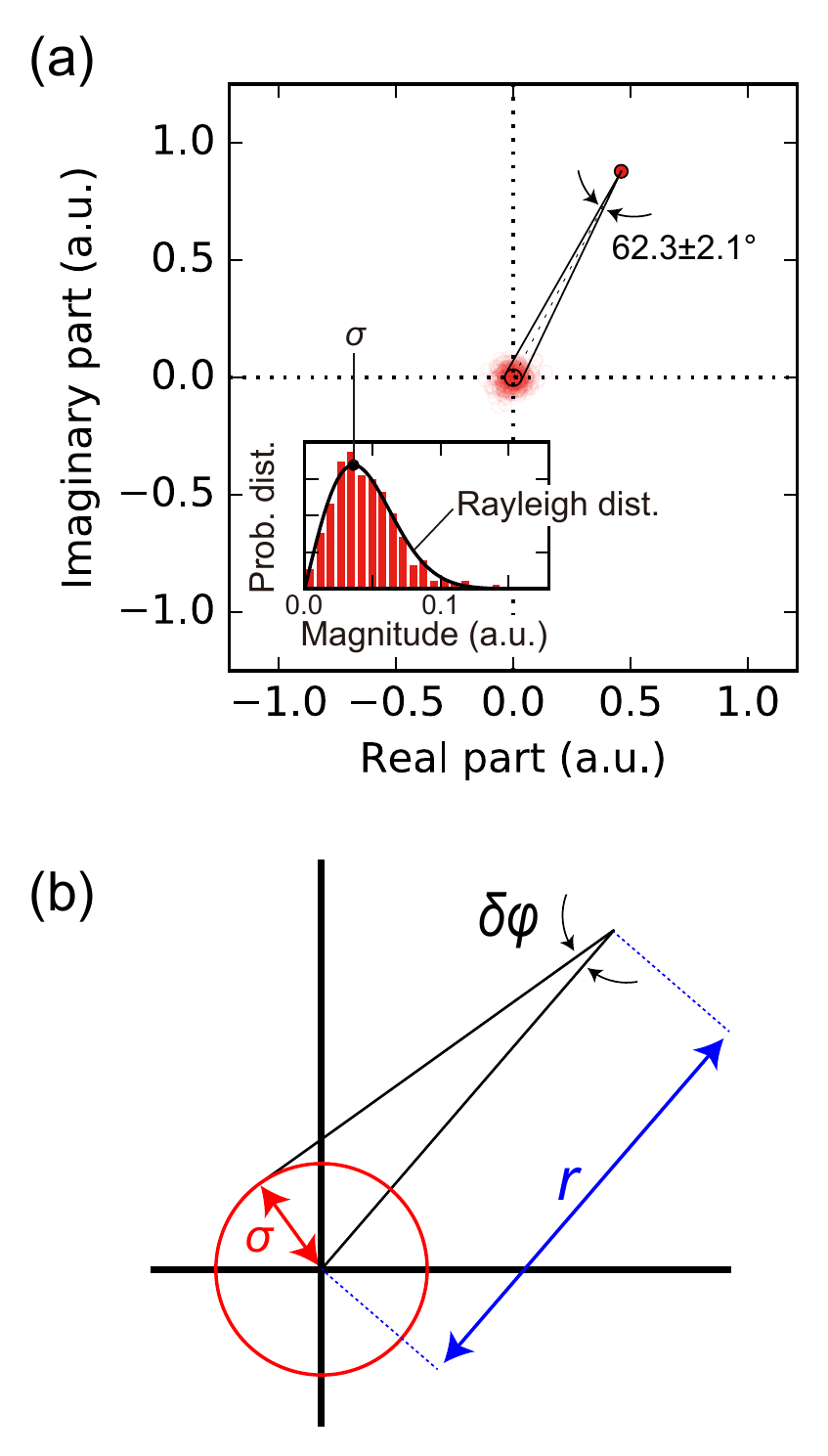}
  \caption{\label{fig4}(a)~Measured Fourier coefficient of the first peak plotted in an IQ diagram, with the inset representing a histogram of noise magnitudes and (b)~definition of the phase precision as a projection angle of noise around the origin from a measured point.}
\end{figure}

\section{\label{sec:principle}Principle}
In this section, we describe our proposed scheme based on iQdyne and compare it with the conventional XY8 measurement.
The iQdyne protocol is a quantum-classical hybrid measurement protocol comprising sequential measurements using DD with a precisely disciplined interval $T_L$ (Fig.~\ref{fig1}(a)).
Splicing the Fourier spectrum at the signal frequency and moving to the IQ diagram, the magnitude of the spectrum depends only on the signal magnitude, and the phase of spectrum directly accords with a signal phase (Fig.~\ref{fig1}(b)).

At first, we formulated the detection of an oscillating magnetic field using XY8 and pointed out inherent challenges. 
Let the magnetic field oscillation obey a cosine, and assuming the time at the origin $t=0$ at the end of the former $\pi$ half pulse, 
\begin{align}
  B_\mathrm{ac}(t) = b_z \cos \left( 2 \pi f_\mathrm{ac} t + \phi_0 \right),
\end{align}
where $b_z$, $\phi_0$, and $f_\mathrm{ac}$ denote the magnitude, initial phase, and frequency of the oscillating field, respectively.
Assuming that the interpulse spacing is half the reciprocal of the signal frequency and ignoring the finite length effect of the MW pulses, the electron spin phase acquired is expressed as
\begin{align}
  \Phi = \frac{2 \gamma_e N_p}{f_\mathrm{ac}} b_z \cos \phi_0 = \kappa b_z \cos \phi_0,
\end{align}
after the XY8 sequence~\cite{DeLange2011,Ishikawa2018}, where $\gamma_e=\SI{28.024e9}{Hz\cdot T^{-1}}$, $N_p$, and $\kappa$ represent the gyromagnetic ratio of the electron spin of the NV center, the number of $\pi$ pulses in an XY8, and the proportional coefficient, respectively.
The spin state is initialized along the $x$-axis of the Bloch sphere and read out by rotating around the $x$-axis, with the resulting fluorescence of XY8 given as
\begin{align}
  S^\mathrm{(XY8)} = C_0 + C \sin \left( \kappa b_z \cos \phi_0 \right),
\end{align}
where $C_0$ and C represent the average count and maximal fluorescence amplitude of the NV center, respectively.
The resulting XY8 sequence signals intricately depend on the magnitude and phase of the oscillating magnetic field.
In a working condition of $\phi_0=0$, it is completely insensitive for a phase shift but most sensitive for magnitude, and vice versa in $\phi_0=\pi/2$.
Therefore, XY8 requires a careful pre-adjustment of the working condition depending on the parameter of interest.
Moreover, if the two parameters change simultaneously, distinguishing their effect becomes challenging.

Next, we show that the dependence of the iQdyne signal on the parameters is simple.
Assuming $T_L$ as an interval between sequential measurements, the oscillating signal phase advances with $\Delta \phi = 2 \pi f_\mathrm{ac} T_L$ from one measurement to the next.
Since the sampling interval $T_L$ is longer than the oscillation period $1/f_\mathrm{ac}$, the resulting signal includes a low-frequency component due to Nyquist folding.
Let $\overline{f_\mathrm{ac}}$ be the apparent frequency of the oscillation, another notation of the advancing phase is
\begin{align}
  \Delta \phi \equiv 2 \pi \overline{f_\mathrm{ac}} T_L \mod 2 \pi.
\end{align}
Then, an initial phase of the oscillating signal of the $m$-th measurement is $\phi_m=\phi_0+m\Delta\phi$.
The iQdyne result of the $m$-th measurement is expressed as
\begin{align}
  S^\mathrm{(iQdyne)}_m &= C_0 + C \sin \left( \kappa b_z \cos \phi_m \right) \\
  &= C_0 + \sum_{n=0}^\infty { I_{2n+1} \cos \left( 2 \pi f_{2n+1} t_m + \theta_{2n+1} \right) }.
\end{align}
Using iQdyne protocol, the oscillating signal at $f_\mathrm{ac}$ appears as several oscillating signals with the apparent frequencies given by $f_{2n+1}= \overline{(2n+1) f_\mathrm{ac}}$, the magnitudes by $I_{2n+1}=2C\cdot\left|J_{2n+1} (\kappa b_z)\right|$, and the initial phases by $|\theta_{2n+1}|=(2n+1)\cdot|\phi_0|$, where $J_{2n+1} (x)$ is the first-kind Bessel function.
These are directly observable using Fourier transform, and the derivation details are provided in the appendix.

Unlike  XY8, the iQdyne protocol involves two output parameters, which the signal magnitude $I_{2n+1}$ depends only on the oscillation magnitude $b_z$, and the phase $\theta_{2n+1}$ depends only on the initial phase $\phi_0$, i.e., two parameters are orthogonal.
The resulting phase $\theta_{2n+1}$ corresponds linearly to the oscillating phase $\phi_0$, and its susceptibility is independent of the oscillation magnitude.
The iQdyne protocol provides a simple and simultaneous measurement of magnitude and phase.
Furthermore, this means that the iQdyne protocol does not need a pre-adjustment of the working condition, since the iQdyne measurement provides the magnitude and phase concurrently, ensuring interpretation is possible under any working condition.

We used an iQdyne protocol on a charge-coupled device (CCD) camera setup.
Each DD measurement of iQdyne was $N_\mathrm{iter}$ times iterated DD sequences and the CCD readout (Fig.~\ref{fig1}(a)), adjusting the iteration period $T_S$ to lock a multiple of the signal period.
This modification enhances sensitivity like the lock-in amplifier and compensates the overhead due to the long readout time of the CCD, but the principle remains unchanged.
We have reported details of this protocol and its frequency characteristics elsewhere~\cite{Mizuno2018}.

\section{\label{sec:responce}Oscillating field measurement}
In this section, demonstration of stroboscopic measurement using iQdyne on a wide-field microscope setup is described.
The sensor involved a shallow and dense NV center ensemble on a IIa $(100)$ diamond substrate (See sample \#1 in the appendix). 
A \SI{532}{nm} laser excitation via epi- and \Koehler-illumination initializes and reads the spin states of the electrons, while a uniform MW radiation via a micro loop coil controls the electron spins. An oscillation signal as a sensing target at \SI{1.908}{MHz} (oscillation period $t_\mathrm{ac}=\SI{524}{ns}$) was generated from a current path created on the surface by photolithography.
Each DD protocol involved the $N_p=64$ pulsed XY8 sequence.
The sensing conditions were as follows: measurement interval $T_S=\SI{23.056}{\us}$, number of iterations $N_\mathrm{iter}=100$, sampling interval $T_L=\SI{7.314850}{ms}$, and measurement length $M=1000$.
Under these conditions, the target signal appears at $\overline{f_\mathrm{ac}}=\SI{49.6}{Hz}$.

The Qdyne response of the first peak magnitude against the applied field magnitude is displayed in Fig.~\ref{fig2}(a)\@.
As derived earlier, this behavior is consistent with the Bessel function $J_1 (\kappa b_z)$, assuming the oscillating field is proportional to an applied voltage.
We swept the oscillation phase (Fig.~\ref{fig2}(b)) at a working condition of  $b_z=\SI{51}{nT}$ ($\kappa b_z=\SI{0.1}{rad}$)\@. 
Since the proportional coefficient between the assigned initial and measured phases is exactly one, this also agrees with the theoretical description.

\section{\label{sec:wf}Wide-field imaging of an oscillating field along a current path}
For stroboscopic imaging of the magnitude and phase of an oscillating magnetic field, we designed another wide-field microscope setup (Fig.~\ref{fig3}(a))\@.
The sensor was a shallow and dense NV center ensemble on a IIa $(111)$ diamond substrate.
The NV center was fabricated by \ce{^{14}N+} ion implantation with \SI{6}{keV} acceleration and \SI{2e13}{{cm}^{-2}} dose, creating an approximately \SI{20}{nm}-thick layer (see sample \#2 in the appendix)\@.
We created a situation with different magnitude and phase of the oscillating field ($f_\mathrm{ac} = \SI{2.0}{MHz}$) in an approximately $100\times\SI{100}{{\micro m}^2}$ observation area.
The oscillating field was generated from a meandering path deposited on the substrate by photolithography (center of Fig.~\ref{fig3}(a))\@.
The fluorescence of the NV center (right of Fig.~\ref{fig3}(a)) corresponds to an intensity distribution of a green laser via a lens array.
The thin sensor layer involved an NV ensemble oriented perpendicular to the substrate surface chosen by selective MW, so that measured signals represent a perpendicular component of the magnetic field vector to the surface.
According to \Ampere's law, the phase measured at the left side of the path should be inverted compared with the right side.
Each DD protocol involved the $N_p=64$ pulsed XY8 sequence, with sensing conditions as follows: measurement interval $T_S=\SI{22.5}{\micro s}$, number of iterations $N_\mathrm{iter}=100$, sampling interval $T_L=\SI{7.259420}{ms}$, and measurement length $M=1000$\@.
A typical spectrum displaying the frequency versus magnitude is shown in Fig.~\ref{fig3}(b), and under the stated conditions, the target signal appears at $\overline{f_\mathrm{ac}}=\SI{22.0}{Hz}$\@.

Maps of the oscillation's magnitude and phase, respectively, are shown in Fig.~\ref{fig3}(c)--(d)\@.
Although magnitude and phase have simultaneously and spatially distributed, our iQdyne protocol can distinguish their changes independently.
Higher magnitudes appear near the current path because the induced magnetic field obeys \Ampere's law.
Comparing the right and left sides of the current path, the measured phases are inverted due to \Ampere's law.
The inside of the meandering corner exhibits constructive interference of the magnetic field, with the zero-magnitude points corresponding to zero of the Bessel function.
This zero line coincides with the boundary of phase inversion.
Beyond the zero line, the phase is also inverted because the strength of the magnitude suffices for the Bessel function $J(\kappa b_z)$ to be minus.

\section{\label{sec:discussion}Discussion}
Considering its application for spectroscopy and testing, small phase shift detection by protocol is crucial.
In this section, we discuss and define the noise floor for phase sensing, namely, precision of the phase shift.
Fig.~\ref{fig4}(a) shows the typical iQdyne spectrum in Fig.~\ref{fig3}(b) plotted on an IQ diagram.
The red point represents the measured Fourier coefficient of the first peak while the semitransparent circles around the origin are Fourier coefficients except for this peak, corresponding to a noise floor.
Assuming the noises contained in the measurements is Gaussian, its distribution on the IQ diagram is a two-dimensional (2D) Gaussian distribution with the origin as the mean.
The probability distribution of the noise's magnitude is a Rayleigh distribution~\cite{Beauchamp1973}, and one of the phases is the uniform distribution on the interval $[0, 2 \pi)$\@.
This forms a circle with radius as the standard deviation of the 2D Gaussian representing the precision of the data point.
Then, we define the precision of the phase as a projected angle of this noise circle viewed from the data point (Fig.~\ref{fig4}(b))\@.
Through a geometric consideration, the precision $\delta \phi$ is formulated as
\begin{align}
  \delta \phi = \sin^{-1} \frac{\sigma}{r},
\end{align}
where $\sigma$ and $r$ are the distance between the origin and the data point and the radius of the noise circle, respectively.
This geometric definition is also justified through statistical estimation theory (see appendix).

For the data in Fig.~\ref{fig4}(a), the phase precision is \SI{2.1}{\degree}, producing a measured phase of $\SI[separate-uncertainty = true]{62.3(21)}{\degree}$, that was acquired at $b_z=\SI{0.76}{\micro T}$, $\kappa b_z=\SI{1.3}{rad}$ in $T_{total} \simeq \SI{6}{min}$ within a $1.2\times\SI{1.2}{{\micro m}^2}$ active area corresponding to a single pixel.
This phase precision means a noise floor of the out-of-phase magnetic field (imaginary part or I-axis) with $b_z \sin \delta \phi = \SI{27}{nT}$\@.

Moreover, we noted the robustness of phase sensing.
The magnitude map (Fig.~\ref{fig3}(c)) was affected by the intensity distribution of laser excitation (Fig.~\ref{fig3}(a), right), because the magnitude corresponded to the fluorescence intensity of NV centers.
However, the phase map (Fig.~\ref{fig3}(d)) indicated no effect from the excitation inhomogeneity.
The reason is that the Fourier transform extracts the phase from the time-development of the fluorescence.
This suggests an interesting approach; the phase sensing by our protocol composes a robust measurement with a signal-to-phase encoding translating some signal of interest as in the oscillation phase.
Developing a concrete method for such encoding is a topic of future work.

\section{\label{sec:coclusion}Conclusions and outlook}
We propose a new measurement protocol for an oscillating magnetic field based on iQdyne.
The approach enables simultaneous detection of magnitude and phase in a wide-field setup without pre-adjusting the measurement condition.
We verified that our protocol provides an orthogonal and readily interpretable measurement of magnitude and phase of the oscillation.
Furthermore, we demonstrated an oscillating magnetic field imaging around a current path involving concurrent distribution of magnitude and phase with an approximately \SI{100}{\micro m} wide-field observation area.
We indicated that the single-pixel phase precision was \SI{2.1}{\degree}, corresponding to a \SI{1.2}{\micro m}-squared region.

The simultaneous measurement of the magnitude and phase of an oscillating magnetic field with high spatial resolution is useful, paving the way to localized current distribution microscopy.
In particular, we note that the high spatial resolution of the NV center is suitable for a phenomenon that is zero in sum, but locally non-zero like a vortex~\cite{Zhou2019}.
Such a technique has significant applications for investigating materials or devices by high-frequency impedance imaging, current distribution on topological or 2D materials, and tiny scale inductive testing, which remain elusive with the existing methods.

%%%% ACKNOWLEDGE %%%
\begin{acknowledgments}
We thank Toshiharu Makino for fabricating the diamond samples.
This work was supported by the MEXT Quantum Leap Flagship Program (MEXT Q-LEAP), Grant Number JPMXS0118067395\@.
\end{acknowledgments}

K.M. conceived the study, built the experimental setup for phase imaging, performed the measurements, analyzed the data, and wrote the manuscript.
H.I. contributed in refining the concept, while Y.M. contributed in building the microwave setup. 
H.I., T.I., and M.H. provided inputs in the discussion section, whereas M.H. supervised the study.

%%%% APENDICES %%%%
\appendix

\section{\label{app:sample}Sample preparation}
We fabricated two high-purity diamond samples named as sample \#1 and \#2\@.
Both samples are single crystalline diamond substrates including natural abundance of \ce{^{13}C} atoms.
Ensemble NV layers were produced by nitrogen ion implantation and subsequent annealing in vacuum, with uniform spatial distribution attributed to the larger ion beam relative to the substrate.
The metal pattern on the surface is a \SI{1}{\micro m}-thick Ti/Cu/Au fabricated by photolithography and electron beam deposition.
The patterns were connected via gold wire bonding.

Sample \#1 has a $(100)$ top facet, with \ce{^{14}N} ions implanted at \SI{6}{keV} energy and \SI{2e13}{{cm}^{-2}} dose at \SI{600}{\degree C}.
Annealing was implemented at \SI{800}{\degree C} for \SI{2}{h}.
The metal pattern includes $\Omega$-shaped MW loop antennas and linear patterns for MW and RF radiation.
Sample \#1 and MW antennas are the same as our previous work~\cite{Mizuno2018}.
The bias magnetic field was around \SI{8}{mT}.

Sample \#2 has $(111)$ top facet, with \ce{^{15}N} ions implanted at \SI{6}{keV} energy and \SI{2e13}{{cm}^{-2}} dose at \SI{600}{\degree C}.
Annealing was implemented at \SI{800}{\degree C} for \SI{2}{h}.
The metal pattern includes meander and linear patterns for RF radiation.
The MW control for the sample \#2 experiment was applied via a planar resonator with the diamond substrate mounted.
This antenna used is a modified version of that reported in our previous work~\cite{Masuyama2018}.
The MW distribution is almost uniform, but the MW strength is slightly stronger at points close to the current path due to coupling between the current path and MW resonator.
The bias magnetic field was around \SI{2}{mT}.

\section{\label{app:setup}Detail of experimental setup}
The excitation light is generated from an optically pumped semiconductor laser (Coherent, Verdi G5) and chopped by an acousto-optic-modulator (Gooch \& Housego, 35250-.2-.53-XQ)\@.
The laser illumination involves epi- and \Koehler-illumination optics for initialization and readout of electron spins.
The laser is focused onto the back focal plane of an objective lens to enlarge the excitation area.
The lens array reduces the spatial coherence and mitigates the interference.
The laser passes through an objective lens and fluorescence from the NV center is collected by the same objective and detected by the EMCCD camera (Andor, iXon3 860)\@.
The objectives are a $60\times$ air objective (Olympus, PlanApoN 60XO) for \#1 and a $20\times$ oil objective (Olympus, MPLFLN20X) for \#2\@.
The MW pulses are generated by a signal generator with a quadrature (IQ) modulation (Keysight, N5182B), switched by Mini-Circuits, ZASWA-2-50DR+, and amplified by R\&K, CGA701M602-4444R\@.
The target signals are generated by a function generator (NF Corporation, WF1967)\@.
All timings are controlled by a data timing generator (Tektronix, DTG5274)\@.
For the Qdyne-type protocol, stabilizing the sampling interval is crucial.
A frequency standard (Stanford Research Systems, FS725) disciplines clocks of all instruments.

\section{\label{app:details_of_derivations}Derivation of dependencies of XY8 and Qdyne on parameters}
In this section, we formulate the dependencies of each measurement protocol on the magnitude $b_z$ and initial phase $\phi_0$.
Let $C$ and $C_0$ be the average fluorescence intensity of an NV center and the amplitude of intensity, respectively, these are then expressed as
\begin{align}
    C &= \frac{{F_{\ket{0}}} - {F_{\ket{1}}}}{2}, \\
    C_0 &= \frac{{F_{\ket{0}}} + {F_{\ket{1}}}}{2},
\end{align}
where $F_{\ket{m_S}}$ is the fluorescence intensity when the spin state of the NV center is $\ket{m_S}$.
We selected the readout axis to be the same direction as the initial superposition state, with the resulting intensity of the XY8 protocol given as
\begin{align}
    S^\mathrm{(XY8)} = C_0 + C \sin \Phi,
\end{align}
where $\Phi$ is the accumulated phase on the spin state~\cite{DeLange2011}.
Assuming an oscillating magnetic field $B_\mathrm{ac} (t)=b_z \cos \left(2 \pi f_\mathrm{ac} t + \phi_0 \right)$, the accumulation by $N_p$-pulsed XY8 sequence is expressed as
\begin{align}
  \Phi &= \left( \int_0^{t_\mathrm{ac}/4} - \int_{t_\mathrm{ac}/4}^{3t_\mathrm{ac}/4} + \cdots \right) 2 \pi \gamma_e B_\mathrm{ac} (t) dt \\
  &= 2 N_p \cdot 2 \pi \gamma_e \int_0^{t_\mathrm{ac}/4} B_\mathrm{ac} (t) dt \\
  &= \frac{2 \gamma_e N_p}{f_\mathrm{ac}} b_z \cos \phi_0 \\
  &= \kappa b_z \cos \phi_0,
\end{align}
with $\kappa$ as the proportional coefficient.
Hence, the resulting intensity of XY8 is given as:
\begin{align}
    S^\mathrm{(XY8)} = C_0 + C \sin \left( \kappa b_z \cos \phi_0 \right).
\end{align}
Their susceptibilities are as follows
\begin{align}
    \frac{\partial S^\mathrm{(XY8)}}{\partial b_z} &= C \cdot \kappa \cdot \cos (\kappa b_z \cos \phi_0) \cdot \cos \phi_0, \\
    \frac{\partial S^\mathrm{(XY8)}}{\partial \phi_0} &= -C \cdot \kappa b_z \cdot \cos (\kappa b_z \cos \phi_0) \cdot \sin \phi_0.
\end{align}

Next, we derive the dependency of Qdyne by assuming $M$ points measurement with a sampling interval $T_L$, with a series of measurements $\{S_m\}_{\left(0\le m < M \right)}$ recorded at timings $t_m = m \cdot T_L$.
Let $\overline{f_\mathrm{ac}}$ be an apparent frequency due to Nyquist folding.
During a sampling interval from $t_m$ to $t_{m+1}$, the oscillating signal phase advances by
\begin{align}
  \Delta \phi = 2 \pi f_\mathrm{ac} T_L.
\end{align}
Then, the initial phase of the oscillating signal of the $m$-th measurement is $\phi_m = \phi_0 + m \Delta \phi \quad (m \ge 0)$.  
The resulting intensity of the $m$-th Qdyne measurement is obtained from
\begin{widetext}
  \begin{align}
      S^\mathrm{(Qdyne)}_m &= C_0 + C \sin (\kappa b_z \cos \phi_m) \\
      &= C_0 + C \sum_{n=0}^\infty (-1)^n \cdot 2 J_{2n+1}(\kappa b_z) \cdot \cos \left( (2n+1) (m \Delta \phi + \phi_0) \right) \\
      &= C_0 + C \sum_{n=0}^\infty (-1)^n \cdot 2 J_{2n+1}(\kappa b_z) \cdot \cos \left( (2n+1) \cdot 2 \pi f_\mathrm{ac} t_m + (2n+1) \phi_0 \right).
  \end{align}
  Here, we used the following relation
  \begin{align}
      \sin (\beta \cos \theta) = \sum_{n=0}^\infty (-1)^n \cdot 2 J_{2n+1}(\beta) \cos \left( (2n+1)\theta \right)).
  \end{align}
  The Qdyne signal is expressed as the sum of odd order components of the signal frequency.
  We can write the $(2n+1)$-th term using the magnitude $I_{2n+1}$, phase $\theta_{2n+1}$, and frequency $f_{2n+1}$ as
  \begin{align}
    (-1)^n \cdot 2 J_{2n+1}(\kappa b_z) \cos\left( 2\pi \cdot (2n+1) f_\mathrm{ac} \cdot t_m + (2n+1) \phi_0 \right) = I_{2n+1} \cos (2 \pi f_{2n+1} t_m + \theta_{2n+1}).
  \end{align}
  Since equivalent parameter sets like $(I,f,\theta)=(I,-f,-\theta)$ exist, we impose constraints $I_{2n+1} \ge 0$ and $f_{2n+1} \ge 0$.
  Therefore, the sign of the phase is determined by the magnetic field $b_z$ and the signal frequency $f_\mathrm{ac}$.
  Alternatively, the sign of $\theta_{2n+1}$ is determined by two factors: the number of times that $\kappa b_z$ crosses zero of the Bessel function and the number of times that the Nyquist folding occurs.
  Eventually, the Qdyne signal is given as follows
  \begin{align}
      S_m^\mathrm{(Qdyne)} = C_0 + \sum_{n=0}^\infty I_{2n+1} \cos(2 \pi f_{2n+1} t_m + \theta_{2n+1}).
  \end{align}
\end{widetext}
So, we obtain $I_{2n+1} = | 2 C J_{2n+1}(\kappa b_z) |$ and $|\theta_{2n+1}|=(2n+1) \cdot |\phi_0|$.
The resulting magnitude $I_{2n+1}$ depends only on the magnetic field $b_z$, whereas the resulting phase $\theta_{2n+1}$ relies only on the initial phase $\phi_0$, i.e., the Qdyne protocol achieves an orthogonalized measurement regarding the magnitude and phase of an oscillating signal.

\section{\label{app:apparent}Nyquist folding and apparent frequency}
In the oversampling condition when the sampling interval $T_L$ is longer than half of the target frequency, the oscillating signal with the frequency $f_\mathrm{ac}$ and initial phase $\phi_0$ resemble another oscillating signal with an apparent frequency $\overline{f_\mathrm{ac}}$ and an apparent phase $\overline{\phi_0}$ due to Nyquist folding.
At first, the phase advancing is congruent with modulo $2 \pi$ with the relationship expressed as:
\begin{align}
    \Delta \phi = 2 \pi T_L f_\mathrm{ac} \equiv 2 \pi T_L \overline{f_\mathrm{ac}} \mod 2 \pi.
\end{align}
In the Fourier spectrum, the frequency axis spans $(-1/2T_L , 1/2T_L]$, but the apparent frequency is in $[0, 1/2T_L]$.
Considering the negative frequency, we obtain a modified phase advancing $\Delta \phi''$ given by
\begin{align}
    \Delta \phi'  &= 2 \pi \left(T_L f_\mathrm{ac} - \lfloor T_L f_\mathrm{ac} \rfloor \right),\\
    \Delta \phi'' &= \min \left\{ \Delta \phi', 2\pi - \Delta\phi' \right\},
\end{align}
where $\lfloor x \rfloor$ is a floor function and $x-\lfloor x \rfloor$ represents the fractional part of $x$.
This produced the apparent frequency as
\begin{align}
    \overline{f_\mathrm{ac}} = \frac{\Delta \phi''}{2 \pi T_L}.
\end{align}
Considering a case where the frequency $f_\mathrm{ac}$ increases from $0$, Nyquist folding occurs under the condition that the product $T_L \times f_\mathrm{ac}$ is an integer or a half-integer.
The apparent phase flips after each folding, and therefore, an apparent phase is obtained from the following
\begin{align}
    \overline{\phi_0} = \phi_0 \cdot (-1)^{\lfloor 2 T_L f_\mathrm{ac} \rfloor}.
\end{align}

\section{\label{app:precision}Definition of phase precision}
In the main text, we define phase precision by a geometric and intuitive illustration.
In this section, we justify this definition through the statistical estimation theory.

Our measurement scheme involves estimation of two parameters, radius $r$ and angle $\theta$ from a complex observed value $Z$ with noise.
Let $X_1$ and $X_2$ be the real and imaginary parts, respectively.
Assuming $X_i\,(i=1,2)$ obey a normal distribution with a variance $\sigma^2$ independently, the probability distribution functions are as follows:
\begin{align}
    f(x_1) &= \frac{1}{\sqrt{2\pi \sigma^2}} \exp\left( -\frac{(x_1 - r\cos\theta)^2}{2\sigma^2} \right), \\
    f(x_2) &= \frac{1}{\sqrt{2\pi \sigma^2}} \exp\left( -\frac{(x_2 - r\sin\theta)^2}{2\sigma^2} \right).
\end{align}
Then, a likelihood function $L(r,\theta)$ and Fisher information matrix $\mathcal{I}$ are obtained as:
\begin{align}
    \ln L(r,\theta) = \sum_{i=1,2} \ln f(x_i),
\end{align}
\begin{align}
    \mathcal{I} = [\partial_i(\ln L) \partial_j(\ln L)]_{i,j \in \{r,\theta\}}
    = \begin{bmatrix}
        1/\sigma^2  &  0  \\
        0  &  r^2 / \sigma^2
    \end{bmatrix}.
  \end{align}
According to the \Cramer--Rao inequality, the standard deviation of any unbiased estimator $\hat{\theta}$ of $\theta$ is bounded by the reciprocal of the square root of the Fisher information, expressed as:
\begin{align}
    \mathrm{std}[\hat{\theta}] \ge \frac{1}{\sqrt{\mathcal{I}_{\theta\theta}}} = \frac{\sigma}{r}.
\end{align}
 
Meanwhile, in this situation, our definition of the phase precision $\delta \theta$ is:
\begin{align}
    \delta \theta = \sin^{-1} \frac{\sigma}{r} \ge \frac{\sigma}{r} = \frac{1}{\sqrt{\mathcal{I}_{\theta\theta}}}.
\end{align}
Therefore, our definition is above the \Cramer--Rao bound and equals the bound asymptotically with a small noise $(\sigma \ll r)$.

%%%% BIBLIOGRAPHY %%%%
\bibliography{main.bib}

\begin{thebibliography}{28}%
\makeatletter
\providecommand \@ifxundefined [1]{%
 \@ifx{#1\undefined}
}%
\providecommand \@ifnum [1]{%
 \ifnum #1\expandafter \@firstoftwo
 \else \expandafter \@secondoftwo
 \fi
}%
\providecommand \@ifx [1]{%
 \ifx #1\expandafter \@firstoftwo
 \else \expandafter \@secondoftwo
 \fi
}%
\providecommand \natexlab [1]{#1}%
\providecommand \enquote  [1]{``#1''}%
\providecommand \bibnamefont  [1]{#1}%
\providecommand \bibfnamefont [1]{#1}%
\providecommand \citenamefont [1]{#1}%
\providecommand \href@noop [0]{\@secondoftwo}%
\providecommand \href [0]{\begingroup \@sanitize@url \@href}%
\providecommand \@href[1]{\@@startlink{#1}\@@href}%
\providecommand \@@href[1]{\endgroup#1\@@endlink}%
\providecommand \@sanitize@url [0]{\catcode `\\12\catcode `\$12\catcode
  `\&12\catcode `\#12\catcode `\^12\catcode `\_12\catcode `\%12\relax}%
\providecommand \@@startlink[1]{}%
\providecommand \@@endlink[0]{}%
\providecommand \url  [0]{\begingroup\@sanitize@url \@url }%
\providecommand \@url [1]{\endgroup\@href {#1}{\urlprefix }}%
\providecommand \urlprefix  [0]{URL }%
\providecommand \Eprint [0]{\href }%
\providecommand \doibase [0]{http://dx.doi.org/}%
\providecommand \selectlanguage [0]{\@gobble}%
\providecommand \bibinfo  [0]{\@secondoftwo}%
\providecommand \bibfield  [0]{\@secondoftwo}%
\providecommand \translation [1]{[#1]}%
\providecommand \BibitemOpen [0]{}%
\providecommand \bibitemStop [0]{}%
\providecommand \bibitemNoStop [0]{.\EOS\space}%
\providecommand \EOS [0]{\spacefactor3000\relax}%
\providecommand \BibitemShut  [1]{\csname bibitem#1\endcsname}%
\let\auto@bib@innerbib\@empty
%</preamble>
\bibitem [{\citenamefont {Degen}\ \emph {et~al.}(2017)\citenamefont {Degen},
  \citenamefont {Reinhard},\ and\ \citenamefont {Cappellaro}}]{Degen2017}%
  \BibitemOpen
  \bibfield  {author} {\bibinfo {author} {\bibfnamefont {C.~L.}\ \bibnamefont
  {Degen}}, \bibinfo {author} {\bibfnamefont {F.}~\bibnamefont {Reinhard}}, \
  and\ \bibinfo {author} {\bibfnamefont {P.}~\bibnamefont {Cappellaro}},\
  }\href {\doibase 10.1103/RevModPhys.89.035002} {\bibfield  {journal}
  {\bibinfo  {journal} {Reviews of Modern Physics}\ }\textbf {\bibinfo {volume}
  {89}},\ \bibinfo {pages} {035002} (\bibinfo {year} {2017})}\BibitemShut
  {NoStop}%
\bibitem [{\citenamefont {Casola}\ \emph {et~al.}(2018)\citenamefont {Casola},
  \citenamefont {van~der Sar},\ and\ \citenamefont {Yacoby}}]{Casola2018}%
  \BibitemOpen
  \bibfield  {author} {\bibinfo {author} {\bibfnamefont {F.}~\bibnamefont
  {Casola}}, \bibinfo {author} {\bibfnamefont {T.}~\bibnamefont {van~der Sar}},
  \ and\ \bibinfo {author} {\bibfnamefont {A.}~\bibnamefont {Yacoby}},\ }\href
  {\doibase 10.1038/natrevmats.2017.88} {\bibfield  {journal} {\bibinfo
  {journal} {Nature Reviews Materials}\ }\textbf {\bibinfo {volume} {3}},\
  \bibinfo {pages} {17088} (\bibinfo {year} {2018})}\BibitemShut {NoStop}%
\bibitem [{\citenamefont {Pezzagna}\ \emph {et~al.}(2010)\citenamefont
  {Pezzagna}, \citenamefont {Naydenov}, \citenamefont {Jelezko}, \citenamefont
  {Wrachtrup},\ and\ \citenamefont {Meijer}}]{Pezzagna2010}%
  \BibitemOpen
  \bibfield  {author} {\bibinfo {author} {\bibfnamefont {S.}~\bibnamefont
  {Pezzagna}}, \bibinfo {author} {\bibfnamefont {B.}~\bibnamefont {Naydenov}},
  \bibinfo {author} {\bibfnamefont {F.}~\bibnamefont {Jelezko}}, \bibinfo
  {author} {\bibfnamefont {J.}~\bibnamefont {Wrachtrup}}, \ and\ \bibinfo
  {author} {\bibfnamefont {J.}~\bibnamefont {Meijer}},\ }\href {\doibase
  10.1088/1367-2630/12/6/065017} {\bibfield  {journal} {\bibinfo  {journal}
  {New Journal of Physics}\ }\textbf {\bibinfo {volume} {12}},\ \bibinfo
  {pages} {065017} (\bibinfo {year} {2010})}\BibitemShut {NoStop}%
\bibitem [{\citenamefont {Ozawa}\ \emph {et~al.}(2019)\citenamefont {Ozawa},
  \citenamefont {Hatano}, \citenamefont {Iwasaki}, \citenamefont {Harada},\
  and\ \citenamefont {Hatano}}]{Ozawa2019}%
  \BibitemOpen
  \bibfield  {author} {\bibinfo {author} {\bibfnamefont {H.}~\bibnamefont
  {Ozawa}}, \bibinfo {author} {\bibfnamefont {Y.}~\bibnamefont {Hatano}},
  \bibinfo {author} {\bibfnamefont {T.}~\bibnamefont {Iwasaki}}, \bibinfo
  {author} {\bibfnamefont {Y.}~\bibnamefont {Harada}}, \ and\ \bibinfo {author}
  {\bibfnamefont {M.}~\bibnamefont {Hatano}},\ }\href {\doibase
  10.7567/1347-4065/ab203c} {\bibfield  {journal} {\bibinfo  {journal}
  {Japanese Journal of Applied Physics}\ }\textbf {\bibinfo {volume} {58}},\
  \bibinfo {pages} {SIIB26} (\bibinfo {year} {2019})}\BibitemShut {NoStop}%
\bibitem [{\citenamefont {Ishiwata}\ \emph {et~al.}(2017)\citenamefont
  {Ishiwata}, \citenamefont {Nakajima}, \citenamefont {Tahara}, \citenamefont
  {Ozawa}, \citenamefont {Iwasaki},\ and\ \citenamefont
  {Hatano}}]{Ishiwata2017}%
  \BibitemOpen
  \bibfield  {author} {\bibinfo {author} {\bibfnamefont {H.}~\bibnamefont
  {Ishiwata}}, \bibinfo {author} {\bibfnamefont {M.}~\bibnamefont {Nakajima}},
  \bibinfo {author} {\bibfnamefont {K.}~\bibnamefont {Tahara}}, \bibinfo
  {author} {\bibfnamefont {H.}~\bibnamefont {Ozawa}}, \bibinfo {author}
  {\bibfnamefont {T.}~\bibnamefont {Iwasaki}}, \ and\ \bibinfo {author}
  {\bibfnamefont {M.}~\bibnamefont {Hatano}},\ }\href {\doibase
  10.1063/1.4993160} {\bibfield  {journal} {\bibinfo  {journal} {Applied
  Physics Letters}\ }\textbf {\bibinfo {volume} {111}},\ \bibinfo {pages}
  {043103} (\bibinfo {year} {2017})}\BibitemShut {NoStop}%
\bibitem [{\citenamefont {Tetienne}\ \emph {et~al.}(2018)\citenamefont
  {Tetienne}, \citenamefont {de~Gille}, \citenamefont {Broadway}, \citenamefont
  {Teraji}, \citenamefont {Lillie}, \citenamefont {McCoey}, \citenamefont
  {Dontschuk}, \citenamefont {Hall}, \citenamefont {Stacey}, \citenamefont
  {Simpson},\ and\ \citenamefont {Hollenberg}}]{Tetienne2017a}%
  \BibitemOpen
  \bibfield  {author} {\bibinfo {author} {\bibfnamefont {J.-P.}\ \bibnamefont
  {Tetienne}}, \bibinfo {author} {\bibfnamefont {R.~W.}\ \bibnamefont
  {de~Gille}}, \bibinfo {author} {\bibfnamefont {D.~A.}\ \bibnamefont
  {Broadway}}, \bibinfo {author} {\bibfnamefont {T.}~\bibnamefont {Teraji}},
  \bibinfo {author} {\bibfnamefont {S.~E.}\ \bibnamefont {Lillie}}, \bibinfo
  {author} {\bibfnamefont {J.~M.}\ \bibnamefont {McCoey}}, \bibinfo {author}
  {\bibfnamefont {N.}~\bibnamefont {Dontschuk}}, \bibinfo {author}
  {\bibfnamefont {L.~T.}\ \bibnamefont {Hall}}, \bibinfo {author}
  {\bibfnamefont {A.}~\bibnamefont {Stacey}}, \bibinfo {author} {\bibfnamefont
  {D.~A.}\ \bibnamefont {Simpson}}, \ and\ \bibinfo {author} {\bibfnamefont
  {L.~C.~L.}\ \bibnamefont {Hollenberg}},\ }\href {\doibase
  10.1103/PhysRevB.97.085402} {\bibfield  {journal} {\bibinfo  {journal}
  {Physical Review B}\ }\textbf {\bibinfo {volume} {97}},\ \bibinfo {pages}
  {085402} (\bibinfo {year} {2018})}\BibitemShut {NoStop}%
\bibitem [{\citenamefont {Ziem}\ \emph {et~al.}(2019)\citenamefont {Ziem},
  \citenamefont {Garsi}, \citenamefont {Fedder},\ and\ \citenamefont
  {Wrachtrup}}]{Ziem2019}%
  \BibitemOpen
  \bibfield  {author} {\bibinfo {author} {\bibfnamefont {F.}~\bibnamefont
  {Ziem}}, \bibinfo {author} {\bibfnamefont {M.}~\bibnamefont {Garsi}},
  \bibinfo {author} {\bibfnamefont {H.}~\bibnamefont {Fedder}}, \ and\ \bibinfo
  {author} {\bibfnamefont {J.}~\bibnamefont {Wrachtrup}},\ }\href {\doibase
  10.1038/s41598-019-47084-w} {\bibfield  {journal} {\bibinfo  {journal}
  {Scientific Reports}\ }\textbf {\bibinfo {volume} {9}},\ \bibinfo {pages}
  {12166} (\bibinfo {year} {2019})}\BibitemShut {NoStop}%
\bibitem [{\citenamefont {Steinert}\ \emph {et~al.}(2013)\citenamefont
  {Steinert}, \citenamefont {Ziem}, \citenamefont {Hall}, \citenamefont
  {Zappe}, \citenamefont {Schweikert}, \citenamefont {G{\"{o}}tz},
  \citenamefont {Aird}, \citenamefont {Balasubramanian}, \citenamefont
  {Hollenberg},\ and\ \citenamefont {Wrachtrup}}]{Steinert2013}%
  \BibitemOpen
  \bibfield  {author} {\bibinfo {author} {\bibfnamefont {S.}~\bibnamefont
  {Steinert}}, \bibinfo {author} {\bibfnamefont {F.}~\bibnamefont {Ziem}},
  \bibinfo {author} {\bibfnamefont {L.~T.}\ \bibnamefont {Hall}}, \bibinfo
  {author} {\bibfnamefont {A.}~\bibnamefont {Zappe}}, \bibinfo {author}
  {\bibfnamefont {M.}~\bibnamefont {Schweikert}}, \bibinfo {author}
  {\bibfnamefont {N.}~\bibnamefont {G{\"{o}}tz}}, \bibinfo {author}
  {\bibfnamefont {A.}~\bibnamefont {Aird}}, \bibinfo {author} {\bibfnamefont
  {G.}~\bibnamefont {Balasubramanian}}, \bibinfo {author} {\bibfnamefont
  {L.}~\bibnamefont {Hollenberg}}, \ and\ \bibinfo {author} {\bibfnamefont
  {J.}~\bibnamefont {Wrachtrup}},\ }\href {\doibase 10.1038/ncomms2588}
  {\bibfield  {journal} {\bibinfo  {journal} {Nature Communications}\ }\textbf
  {\bibinfo {volume} {4}},\ \bibinfo {pages} {1607} (\bibinfo {year}
  {2013})}\BibitemShut {NoStop}%
\bibitem [{\citenamefont {Horsley}\ \emph {et~al.}(2018)\citenamefont
  {Horsley}, \citenamefont {Appel}, \citenamefont {Wolters}, \citenamefont
  {Achard}, \citenamefont {Tallaire}, \citenamefont {Maletinsky},\ and\
  \citenamefont {Treutlein}}]{Horsley2018}%
  \BibitemOpen
  \bibfield  {author} {\bibinfo {author} {\bibfnamefont {A.}~\bibnamefont
  {Horsley}}, \bibinfo {author} {\bibfnamefont {P.}~\bibnamefont {Appel}},
  \bibinfo {author} {\bibfnamefont {J.}~\bibnamefont {Wolters}}, \bibinfo
  {author} {\bibfnamefont {J.}~\bibnamefont {Achard}}, \bibinfo {author}
  {\bibfnamefont {A.}~\bibnamefont {Tallaire}}, \bibinfo {author}
  {\bibfnamefont {P.}~\bibnamefont {Maletinsky}}, \ and\ \bibinfo {author}
  {\bibfnamefont {P.}~\bibnamefont {Treutlein}},\ }\href {\doibase
  10.1103/PhysRevApplied.10.044039} {\bibfield  {journal} {\bibinfo  {journal}
  {Physical Review Applied}\ }\textbf {\bibinfo {volume} {10}},\ \bibinfo
  {pages} {044039} (\bibinfo {year} {2018})}\BibitemShut {NoStop}%
\bibitem [{\citenamefont {Fescenko}\ \emph {et~al.}(2019)\citenamefont
  {Fescenko}, \citenamefont {Laraoui}, \citenamefont {Smits}, \citenamefont
  {Mosavian}, \citenamefont {Kehayias}, \citenamefont {Seto}, \citenamefont
  {Bougas}, \citenamefont {Jarmola},\ and\ \citenamefont
  {Acosta}}]{Fescenko2018}%
  \BibitemOpen
  \bibfield  {author} {\bibinfo {author} {\bibfnamefont {I.}~\bibnamefont
  {Fescenko}}, \bibinfo {author} {\bibfnamefont {A.}~\bibnamefont {Laraoui}},
  \bibinfo {author} {\bibfnamefont {J.}~\bibnamefont {Smits}}, \bibinfo
  {author} {\bibfnamefont {N.}~\bibnamefont {Mosavian}}, \bibinfo {author}
  {\bibfnamefont {P.}~\bibnamefont {Kehayias}}, \bibinfo {author}
  {\bibfnamefont {J.}~\bibnamefont {Seto}}, \bibinfo {author} {\bibfnamefont
  {L.}~\bibnamefont {Bougas}}, \bibinfo {author} {\bibfnamefont
  {A.}~\bibnamefont {Jarmola}}, \ and\ \bibinfo {author} {\bibfnamefont
  {V.~M.}\ \bibnamefont {Acosta}},\ }\href {\doibase
  10.1103/PhysRevApplied.11.034029} {\bibfield  {journal} {\bibinfo  {journal}
  {Physical Review Applied}\ }\textbf {\bibinfo {volume} {11}},\ \bibinfo
  {pages} {034029} (\bibinfo {year} {2019})}\BibitemShut {NoStop}%
\bibitem [{\citenamefont {Tetienne}\ \emph {et~al.}(2017)\citenamefont
  {Tetienne}, \citenamefont {Dontschuk}, \citenamefont {Broadway},
  \citenamefont {Stacey}, \citenamefont {Simpson},\ and\ \citenamefont
  {Hollenberg}}]{Tetienne2017}%
  \BibitemOpen
  \bibfield  {author} {\bibinfo {author} {\bibfnamefont {J.-P.}\ \bibnamefont
  {Tetienne}}, \bibinfo {author} {\bibfnamefont {N.}~\bibnamefont {Dontschuk}},
  \bibinfo {author} {\bibfnamefont {D.~A.}\ \bibnamefont {Broadway}}, \bibinfo
  {author} {\bibfnamefont {A.}~\bibnamefont {Stacey}}, \bibinfo {author}
  {\bibfnamefont {D.~A.}\ \bibnamefont {Simpson}}, \ and\ \bibinfo {author}
  {\bibfnamefont {L.~C.~L.}\ \bibnamefont {Hollenberg}},\ }\href {\doibase
  10.1126/sciadv.1602429} {\bibfield  {journal} {\bibinfo  {journal} {Science
  Advances}\ }\textbf {\bibinfo {volume} {3}},\ \bibinfo {pages} {e1602429}
  (\bibinfo {year} {2017})}\BibitemShut {NoStop}%
\bibitem [{\citenamefont {Schlussel}\ \emph {et~al.}(2018)\citenamefont
  {Schlussel}, \citenamefont {Lenz}, \citenamefont {Rohner}, \citenamefont
  {Bar-Haim}, \citenamefont {Bougas}, \citenamefont {Groswasser}, \citenamefont
  {Kieschnick}, \citenamefont {Rozenberg}, \citenamefont {Thiel}, \citenamefont
  {Waxman}, \citenamefont {Meijer}, \citenamefont {Maletinsky}, \citenamefont
  {Budker},\ and\ \citenamefont {Folman}}]{Schlussel2018}%
  \BibitemOpen
  \bibfield  {author} {\bibinfo {author} {\bibfnamefont {Y.}~\bibnamefont
  {Schlussel}}, \bibinfo {author} {\bibfnamefont {T.}~\bibnamefont {Lenz}},
  \bibinfo {author} {\bibfnamefont {D.}~\bibnamefont {Rohner}}, \bibinfo
  {author} {\bibfnamefont {Y.}~\bibnamefont {Bar-Haim}}, \bibinfo {author}
  {\bibfnamefont {L.}~\bibnamefont {Bougas}}, \bibinfo {author} {\bibfnamefont
  {D.}~\bibnamefont {Groswasser}}, \bibinfo {author} {\bibfnamefont
  {M.}~\bibnamefont {Kieschnick}}, \bibinfo {author} {\bibfnamefont
  {E.}~\bibnamefont {Rozenberg}}, \bibinfo {author} {\bibfnamefont
  {L.}~\bibnamefont {Thiel}}, \bibinfo {author} {\bibfnamefont
  {A.}~\bibnamefont {Waxman}}, \bibinfo {author} {\bibfnamefont
  {J.}~\bibnamefont {Meijer}}, \bibinfo {author} {\bibfnamefont
  {P.}~\bibnamefont {Maletinsky}}, \bibinfo {author} {\bibfnamefont
  {D.}~\bibnamefont {Budker}}, \ and\ \bibinfo {author} {\bibfnamefont
  {R.}~\bibnamefont {Folman}},\ }\href {\doibase
  10.1103/PhysRevApplied.10.034032} {\bibfield  {journal} {\bibinfo  {journal}
  {Physical Review Applied}\ }\textbf {\bibinfo {volume} {10}},\ \bibinfo
  {pages} {034032} (\bibinfo {year} {2018})}\BibitemShut {NoStop}%
\bibitem [{\citenamefont {Lillie}\ \emph {et~al.}(2019)\citenamefont {Lillie},
  \citenamefont {Dontschuk}, \citenamefont {Broadway}, \citenamefont {Creedon},
  \citenamefont {Hollenberg},\ and\ \citenamefont {Tetienne}}]{Lillie2019}%
  \BibitemOpen
  \bibfield  {author} {\bibinfo {author} {\bibfnamefont {S.~E.}\ \bibnamefont
  {Lillie}}, \bibinfo {author} {\bibfnamefont {N.}~\bibnamefont {Dontschuk}},
  \bibinfo {author} {\bibfnamefont {D.~A.}\ \bibnamefont {Broadway}}, \bibinfo
  {author} {\bibfnamefont {D.~L.}\ \bibnamefont {Creedon}}, \bibinfo {author}
  {\bibfnamefont {L.~C.}\ \bibnamefont {Hollenberg}}, \ and\ \bibinfo {author}
  {\bibfnamefont {J.-P.}\ \bibnamefont {Tetienne}},\ }\href {\doibase
  10.1103/PhysRevApplied.12.024018} {\bibfield  {journal} {\bibinfo  {journal}
  {Physical Review Applied}\ }\textbf {\bibinfo {volume} {12}},\ \bibinfo
  {pages} {024018} (\bibinfo {year} {2019})}\BibitemShut {NoStop}%
\bibitem [{\citenamefont {Tetienne}\ \emph {et~al.}(2019)\citenamefont
  {Tetienne}, \citenamefont {Dontschuk}, \citenamefont {Broadway},
  \citenamefont {Lillie}, \citenamefont {Teraji}, \citenamefont {Simpson},
  \citenamefont {Stacey},\ and\ \citenamefont {Hollenberg}}]{Tetienne2019}%
  \BibitemOpen
  \bibfield  {author} {\bibinfo {author} {\bibfnamefont {J.-P.}\ \bibnamefont
  {Tetienne}}, \bibinfo {author} {\bibfnamefont {N.}~\bibnamefont {Dontschuk}},
  \bibinfo {author} {\bibfnamefont {D.~A.}\ \bibnamefont {Broadway}}, \bibinfo
  {author} {\bibfnamefont {S.~E.}\ \bibnamefont {Lillie}}, \bibinfo {author}
  {\bibfnamefont {T.}~\bibnamefont {Teraji}}, \bibinfo {author} {\bibfnamefont
  {D.~A.}\ \bibnamefont {Simpson}}, \bibinfo {author} {\bibfnamefont
  {A.}~\bibnamefont {Stacey}}, \ and\ \bibinfo {author} {\bibfnamefont
  {L.~C.~L.}\ \bibnamefont {Hollenberg}},\ }\href {\doibase
  10.1103/PhysRevB.99.014436} {\bibfield  {journal} {\bibinfo  {journal}
  {Physical Review B}\ }\textbf {\bibinfo {volume} {99}},\ \bibinfo {pages}
  {014436} (\bibinfo {year} {2019})}\BibitemShut {NoStop}%
\bibitem [{\citenamefont {Zhou}\ \emph {et~al.}(2020)\citenamefont {Zhou},
  \citenamefont {Jerger}, \citenamefont {Lee}, \citenamefont {Fukami},
  \citenamefont {Mujid}, \citenamefont {Park},\ and\ \citenamefont
  {Awschalom}}]{Zhou2019}%
  \BibitemOpen
  \bibfield  {author} {\bibinfo {author} {\bibfnamefont {B.~B.}\ \bibnamefont
  {Zhou}}, \bibinfo {author} {\bibfnamefont {P.~C.}\ \bibnamefont {Jerger}},
  \bibinfo {author} {\bibfnamefont {K.-H.}\ \bibnamefont {Lee}}, \bibinfo
  {author} {\bibfnamefont {M.}~\bibnamefont {Fukami}}, \bibinfo {author}
  {\bibfnamefont {F.}~\bibnamefont {Mujid}}, \bibinfo {author} {\bibfnamefont
  {J.}~\bibnamefont {Park}}, \ and\ \bibinfo {author} {\bibfnamefont {D.~D.}\
  \bibnamefont {Awschalom}},\ }\href {\doibase 10.1103/PhysRevX.10.011003}
  {\bibfield  {journal} {\bibinfo  {journal} {Physical Review X}\ }\textbf
  {\bibinfo {volume} {10}},\ \bibinfo {pages} {011003} (\bibinfo {year}
  {2020})}\BibitemShut {NoStop}%
\bibitem [{\citenamefont {Chatzidrosos}\ \emph {et~al.}(2019)\citenamefont
  {Chatzidrosos}, \citenamefont {Wickenbrock}, \citenamefont {Bougas},
  \citenamefont {Zheng}, \citenamefont {Tretiak}, \citenamefont {Yang},\ and\
  \citenamefont {Budker}}]{Chatzidrosos2019}%
  \BibitemOpen
  \bibfield  {author} {\bibinfo {author} {\bibfnamefont {G.}~\bibnamefont
  {Chatzidrosos}}, \bibinfo {author} {\bibfnamefont {A.}~\bibnamefont
  {Wickenbrock}}, \bibinfo {author} {\bibfnamefont {L.}~\bibnamefont {Bougas}},
  \bibinfo {author} {\bibfnamefont {H.}~\bibnamefont {Zheng}}, \bibinfo
  {author} {\bibfnamefont {O.}~\bibnamefont {Tretiak}}, \bibinfo {author}
  {\bibfnamefont {Y.}~\bibnamefont {Yang}}, \ and\ \bibinfo {author}
  {\bibfnamefont {D.}~\bibnamefont {Budker}},\ }\href {\doibase
  10.1103/PhysRevApplied.11.014060} {\bibfield  {journal} {\bibinfo  {journal}
  {Physical Review Applied}\ }\textbf {\bibinfo {volume} {11}},\ \bibinfo
  {pages} {014060} (\bibinfo {year} {2019})}\BibitemShut {NoStop}%
\bibitem [{\citenamefont {Flanagan}\ \emph {et~al.}(1990)\citenamefont
  {Flanagan}, \citenamefont {Jordan},\ and\ \citenamefont
  {Whittington}}]{Flanagan1990}%
  \BibitemOpen
  \bibfield  {author} {\bibinfo {author} {\bibfnamefont {I.~M.}\ \bibnamefont
  {Flanagan}}, \bibinfo {author} {\bibfnamefont {J.~R.}\ \bibnamefont
  {Jordan}}, \ and\ \bibinfo {author} {\bibfnamefont {H.~W.}\ \bibnamefont
  {Whittington}},\ }\href {\doibase 10.1088/0957-0233/1/5/001} {\bibfield
  {journal} {\bibinfo  {journal} {Measurement Science and Technology}\ }\textbf
  {\bibinfo {volume} {1}},\ \bibinfo {pages} {381} (\bibinfo {year}
  {1990})}\BibitemShut {NoStop}%
\bibitem [{\citenamefont {Ribeiro}\ and\ \citenamefont
  {Ramos}(2008)}]{Ribeiro2008}%
  \BibitemOpen
  \bibfield  {author} {\bibinfo {author} {\bibfnamefont {A.~L.}\ \bibnamefont
  {Ribeiro}}\ and\ \bibinfo {author} {\bibfnamefont {H.~G.}\ \bibnamefont
  {Ramos}},\ }in\ \href {\doibase 10.1109/IMTC.2008.4547270} {\emph {\bibinfo
  {booktitle} {2008 IEEE Instrumentation and Measurement Technology
  Conference}}}\ (\bibinfo  {publisher} {IEEE},\ \bibinfo {year} {2008})\ p.\
  \bibinfo {pages} {1447}\BibitemShut {NoStop}%
\bibitem [{\citenamefont {Wolf}\ \emph {et~al.}(2015)\citenamefont {Wolf},
  \citenamefont {Neumann}, \citenamefont {Nakamura}, \citenamefont {Sumiya},
  \citenamefont {Ohshima}, \citenamefont {Isoya},\ and\ \citenamefont
  {Wrachtrup}}]{Wolf2015}%
  \BibitemOpen
  \bibfield  {author} {\bibinfo {author} {\bibfnamefont {T.}~\bibnamefont
  {Wolf}}, \bibinfo {author} {\bibfnamefont {P.}~\bibnamefont {Neumann}},
  \bibinfo {author} {\bibfnamefont {K.}~\bibnamefont {Nakamura}}, \bibinfo
  {author} {\bibfnamefont {H.}~\bibnamefont {Sumiya}}, \bibinfo {author}
  {\bibfnamefont {T.}~\bibnamefont {Ohshima}}, \bibinfo {author} {\bibfnamefont
  {J.}~\bibnamefont {Isoya}}, \ and\ \bibinfo {author} {\bibfnamefont
  {J.}~\bibnamefont {Wrachtrup}},\ }\href {\doibase 10.1103/PhysRevX.5.041001}
  {\bibfield  {journal} {\bibinfo  {journal} {Physical Review X}\ }\textbf
  {\bibinfo {volume} {5}},\ \bibinfo {pages} {041001} (\bibinfo {year}
  {2015})}\BibitemShut {NoStop}%
\bibitem [{\citenamefont {Barry}\ \emph {et~al.}(2016)\citenamefont {Barry},
  \citenamefont {Turner}, \citenamefont {Schloss}, \citenamefont {Glenn},
  \citenamefont {Song}, \citenamefont {Lukin}, \citenamefont {Park},\ and\
  \citenamefont {Walsworth}}]{Barry2016}%
  \BibitemOpen
  \bibfield  {author} {\bibinfo {author} {\bibfnamefont {J.~F.}\ \bibnamefont
  {Barry}}, \bibinfo {author} {\bibfnamefont {M.~J.}\ \bibnamefont {Turner}},
  \bibinfo {author} {\bibfnamefont {J.~M.}\ \bibnamefont {Schloss}}, \bibinfo
  {author} {\bibfnamefont {D.~R.}\ \bibnamefont {Glenn}}, \bibinfo {author}
  {\bibfnamefont {Y.}~\bibnamefont {Song}}, \bibinfo {author} {\bibfnamefont
  {M.~D.}\ \bibnamefont {Lukin}}, \bibinfo {author} {\bibfnamefont
  {H.}~\bibnamefont {Park}}, \ and\ \bibinfo {author} {\bibfnamefont {R.~L.}\
  \bibnamefont {Walsworth}},\ }\href {\doibase 10.1073/pnas.1601513113}
  {\bibfield  {journal} {\bibinfo  {journal} {Proceedings of the National
  Academy of Sciences}\ }\textbf {\bibinfo {volume} {113}},\ \bibinfo {pages}
  {14133} (\bibinfo {year} {2016})}\BibitemShut {NoStop}%
\bibitem [{\citenamefont {Masuyama}\ \emph {et~al.}(2018)\citenamefont
  {Masuyama}, \citenamefont {Mizuno}, \citenamefont {Ozawa}, \citenamefont
  {Ishiwata}, \citenamefont {Hatano}, \citenamefont {Ohshima}, \citenamefont
  {Iwasaki},\ and\ \citenamefont {Hatano}}]{Masuyama2018}%
  \BibitemOpen
  \bibfield  {author} {\bibinfo {author} {\bibfnamefont {Y.}~\bibnamefont
  {Masuyama}}, \bibinfo {author} {\bibfnamefont {K.}~\bibnamefont {Mizuno}},
  \bibinfo {author} {\bibfnamefont {H.}~\bibnamefont {Ozawa}}, \bibinfo
  {author} {\bibfnamefont {H.}~\bibnamefont {Ishiwata}}, \bibinfo {author}
  {\bibfnamefont {Y.}~\bibnamefont {Hatano}}, \bibinfo {author} {\bibfnamefont
  {T.}~\bibnamefont {Ohshima}}, \bibinfo {author} {\bibfnamefont
  {T.}~\bibnamefont {Iwasaki}}, \ and\ \bibinfo {author} {\bibfnamefont
  {M.}~\bibnamefont {Hatano}},\ }\href {\doibase 10.1063/1.5047078} {\bibfield
  {journal} {\bibinfo  {journal} {Review of Scientific Instruments}\ }\textbf
  {\bibinfo {volume} {89}},\ \bibinfo {pages} {125007} (\bibinfo {year}
  {2018})}\BibitemShut {NoStop}%
\bibitem [{\citenamefont {Ishikawa}\ \emph {et~al.}(2019)\citenamefont
  {Ishikawa}, \citenamefont {Yoshizawa}, \citenamefont {Mawatari},
  \citenamefont {Kashiwaya},\ and\ \citenamefont {Watanabe}}]{Ishikawa2019a}%
  \BibitemOpen
  \bibfield  {author} {\bibinfo {author} {\bibfnamefont {T.}~\bibnamefont
  {Ishikawa}}, \bibinfo {author} {\bibfnamefont {A.}~\bibnamefont {Yoshizawa}},
  \bibinfo {author} {\bibfnamefont {Y.}~\bibnamefont {Mawatari}}, \bibinfo
  {author} {\bibfnamefont {S.}~\bibnamefont {Kashiwaya}}, \ and\ \bibinfo
  {author} {\bibfnamefont {H.}~\bibnamefont {Watanabe}},\ }\href {\doibase
  10.1063/1.5096610} {\bibfield  {journal} {\bibinfo  {journal} {Journal of
  Applied Physics}\ }\textbf {\bibinfo {volume} {126}},\ \bibinfo {pages}
  {064504} (\bibinfo {year} {2019})}\BibitemShut {NoStop}%
\bibitem [{\citenamefont {de~Lange}\ \emph {et~al.}(2011)\citenamefont
  {de~Lange}, \citenamefont {Rist{\`{e}}}, \citenamefont {Dobrovitski},\ and\
  \citenamefont {Hanson}}]{DeLange2011}%
  \BibitemOpen
  \bibfield  {author} {\bibinfo {author} {\bibfnamefont {G.}~\bibnamefont
  {de~Lange}}, \bibinfo {author} {\bibfnamefont {D.}~\bibnamefont
  {Rist{\`{e}}}}, \bibinfo {author} {\bibfnamefont {V.~V.}\ \bibnamefont
  {Dobrovitski}}, \ and\ \bibinfo {author} {\bibfnamefont {R.}~\bibnamefont
  {Hanson}},\ }\href {\doibase 10.1103/PhysRevLett.106.080802} {\bibfield
  {journal} {\bibinfo  {journal} {Physical Review Letters}\ }\textbf {\bibinfo
  {volume} {106}},\ \bibinfo {pages} {080802} (\bibinfo {year}
  {2011})}\BibitemShut {NoStop}%
\bibitem [{\citenamefont {Mizuno}\ \emph {et~al.}(2018)\citenamefont {Mizuno},
  \citenamefont {Nakajima}, \citenamefont {Ishiwata}, \citenamefont {Masuyama},
  \citenamefont {Iwasaki},\ and\ \citenamefont {Hatano}}]{Mizuno2018}%
  \BibitemOpen
  \bibfield  {author} {\bibinfo {author} {\bibfnamefont {K.}~\bibnamefont
  {Mizuno}}, \bibinfo {author} {\bibfnamefont {M.}~\bibnamefont {Nakajima}},
  \bibinfo {author} {\bibfnamefont {H.}~\bibnamefont {Ishiwata}}, \bibinfo
  {author} {\bibfnamefont {Y.}~\bibnamefont {Masuyama}}, \bibinfo {author}
  {\bibfnamefont {T.}~\bibnamefont {Iwasaki}}, \ and\ \bibinfo {author}
  {\bibfnamefont {M.}~\bibnamefont {Hatano}},\ }\href {\doibase
  10.1063/1.5048265} {\bibfield  {journal} {\bibinfo  {journal} {AIP Advances}\
  }\textbf {\bibinfo {volume} {8}},\ \bibinfo {pages} {125316} (\bibinfo {year}
  {2018})}\BibitemShut {NoStop}%
\bibitem [{\citenamefont {Schmitt}\ \emph {et~al.}(2017)\citenamefont
  {Schmitt}, \citenamefont {Gefen}, \citenamefont {St{\"{u}}rner},
  \citenamefont {Unden}, \citenamefont {Wolff}, \citenamefont {M{\"{u}}ller},
  \citenamefont {Scheuer}, \citenamefont {Naydenov}, \citenamefont {Markham},
  \citenamefont {Pezzagna}, \citenamefont {Meijer}, \citenamefont {Schwarz},
  \citenamefont {Plenio}, \citenamefont {Retzker}, \citenamefont {McGuinness},\
  and\ \citenamefont {Jelezko}}]{Schmitt2017}%
  \BibitemOpen
  \bibfield  {author} {\bibinfo {author} {\bibfnamefont {S.}~\bibnamefont
  {Schmitt}}, \bibinfo {author} {\bibfnamefont {T.}~\bibnamefont {Gefen}},
  \bibinfo {author} {\bibfnamefont {F.~M.}\ \bibnamefont {St{\"{u}}rner}},
  \bibinfo {author} {\bibfnamefont {T.}~\bibnamefont {Unden}}, \bibinfo
  {author} {\bibfnamefont {G.}~\bibnamefont {Wolff}}, \bibinfo {author}
  {\bibfnamefont {C.}~\bibnamefont {M{\"{u}}ller}}, \bibinfo {author}
  {\bibfnamefont {J.}~\bibnamefont {Scheuer}}, \bibinfo {author} {\bibfnamefont
  {B.}~\bibnamefont {Naydenov}}, \bibinfo {author} {\bibfnamefont
  {M.}~\bibnamefont {Markham}}, \bibinfo {author} {\bibfnamefont
  {S.}~\bibnamefont {Pezzagna}}, \bibinfo {author} {\bibfnamefont
  {J.}~\bibnamefont {Meijer}}, \bibinfo {author} {\bibfnamefont
  {I.}~\bibnamefont {Schwarz}}, \bibinfo {author} {\bibfnamefont
  {M.}~\bibnamefont {Plenio}}, \bibinfo {author} {\bibfnamefont
  {A.}~\bibnamefont {Retzker}}, \bibinfo {author} {\bibfnamefont {L.~P.}\
  \bibnamefont {McGuinness}}, \ and\ \bibinfo {author} {\bibfnamefont
  {F.}~\bibnamefont {Jelezko}},\ }\href {\doibase 10.1126/science.aam5532}
  {\bibfield  {journal} {\bibinfo  {journal} {Science}\ }\textbf {\bibinfo
  {volume} {356}},\ \bibinfo {pages} {832} (\bibinfo {year}
  {2017})}\BibitemShut {NoStop}%
\bibitem [{\citenamefont {Boss}\ \emph {et~al.}(2017)\citenamefont {Boss},
  \citenamefont {Cujia}, \citenamefont {Zopes},\ and\ \citenamefont
  {Degen}}]{Boss2017}%
  \BibitemOpen
  \bibfield  {author} {\bibinfo {author} {\bibfnamefont {J.~M.}\ \bibnamefont
  {Boss}}, \bibinfo {author} {\bibfnamefont {K.~S.}\ \bibnamefont {Cujia}},
  \bibinfo {author} {\bibfnamefont {J.}~\bibnamefont {Zopes}}, \ and\ \bibinfo
  {author} {\bibfnamefont {C.~L.}\ \bibnamefont {Degen}},\ }\href {\doibase
  10.1126/science.aam7009} {\bibfield  {journal} {\bibinfo  {journal}
  {Science}\ }\textbf {\bibinfo {volume} {356}},\ \bibinfo {pages} {837}
  (\bibinfo {year} {2017})}\BibitemShut {NoStop}%
\bibitem [{\citenamefont {Ishikawa}\ \emph {et~al.}(2018)\citenamefont
  {Ishikawa}, \citenamefont {Yoshizwa}, \citenamefont {Mawatari}, \citenamefont
  {Kashiwaya},\ and\ \citenamefont {Watanabe}}]{Ishikawa2018}%
  \BibitemOpen
  \bibfield  {author} {\bibinfo {author} {\bibfnamefont {T.}~\bibnamefont
  {Ishikawa}}, \bibinfo {author} {\bibfnamefont {A.}~\bibnamefont {Yoshizwa}},
  \bibinfo {author} {\bibfnamefont {Y.}~\bibnamefont {Mawatari}}, \bibinfo
  {author} {\bibfnamefont {S.}~\bibnamefont {Kashiwaya}}, \ and\ \bibinfo
  {author} {\bibfnamefont {H.}~\bibnamefont {Watanabe}},\ }\href {\doibase
  10.1103/PhysRevApplied.10.054059} {\bibfield  {journal} {\bibinfo  {journal}
  {Physical Review Applied}\ }\textbf {\bibinfo {volume} {10}},\ \bibinfo
  {pages} {1} (\bibinfo {year} {2018})}\BibitemShut {NoStop}%
\bibitem [{\citenamefont {Beauchamp}(1973)}]{Beauchamp1973}%
  \BibitemOpen
  \bibfield  {author} {\bibinfo {author} {\bibfnamefont {K.~G.}\ \bibnamefont
  {Beauchamp}},\ }\href@noop {} {\emph {\bibinfo {title} {{Signal processing:
  using analog and digital techniques}}}}\ (\bibinfo  {publisher} {Allen {\&}
  Unwin},\ \bibinfo {address} {London},\ \bibinfo {year} {1973})\BibitemShut
  {NoStop}%
\end{thebibliography}%
%\clearpage

\end{document}